\documentclass[aps,prl,twocolumn,preprintnumbers,groupedaddress,superscriptaddress,floatfix,tightenlines,reprint]{revtex4-1}
\usepackage{mathrsfs,natbib}
\usepackage{graphics,epsfig,color, graphicx}
\usepackage{verbatim}
\usepackage{bm}

\usepackage{graphicx,epsf,amssymb,bbm,amsbsy,amsfonts,amssymb,amsmath}
\usepackage{hyperref}

\newcommand{\eq}{\begin{equation}}
\newcommand{\eqe}{\end{equation}}

\def\tr{\text{tr}\,}

\newcommand{\eqa}{\begin{eqnarray}}
\newcommand{\eqae}{\end{eqnarray}}

\def\Z{{\mathbb Z}}
\def\R{{\mathbb R}}

\usepackage{verbatim}   
\usepackage[all]{xy}
\usepackage{bm}  
\def\Dslash{{\rlap{\raise 1pt \hbox{$\>/$}}D}}
\def\Pslash{{\rlap{\raise  1pt \hbox{$\>/$}}\,\partial}}

\hypersetup{
    colorlinks=true,       
    linkcolor=red,          
    citecolor=blue,        
    filecolor=magenta,      
    urlcolor=blue           
}

\begin{document}

\title{Chiral Lagrangian from   Duality and  Monopole Operators  in Compactified 
QCD}

\preprint{INT-PUB-16-011}

\author{Aleksey Cherman}
\email{aleksey.cherman.physics@gmail.com}
\affiliation{Institute for Nuclear Theory, University of Washington, Seattle, WA 98105 USA}

\author{Thomas Sch\"afer}
\email{tmschaef@ncsu.edu}
\affiliation{Department of Physics, North Carolina State University, Raleigh, NC 27695, USA}

\author{Mithat \"Unsal}
\email{unsal.mithat@gmail.com}
\affiliation{Department of Physics, North Carolina State University, Raleigh, NC 27695, USA}

\begin{abstract}
We show that there exists a special compactification of QCD on $\mathbb{R}^3 \times S^1$ in which the theory has a domain where continuous chiral symmetry breaking is analytically  calculable. 
  We  give a microscopic derivation of the chiral lagrangian, the chiral 
condensate, and the Gell-Mann-Oakes-Renner relation $m_{\pi}^2 f_{\pi}^2 
= m_q \langle \bar{q} q \rangle$.  Abelian duality,  monopole operators,  and 
 flavor-twisted boundary conditions 
play the main roles.  The flavor twisting leads to 
the  new effect of  fractional jumping of fermion zero modes among monopole-instantons. 
 Chiral symmetry breaking
is induced by monopole-instanton operators, and the Nambu-Goldstone pions 
arise by color-flavor transmutation from gapless ``dual photons". 
We also give a microscopic picture 
of the ``constituent quark" masses.  
Our 
results are consistent with expectations from chiral perturbation theory 
at large $S^1$, and yield strong support for adiabatic continuity between the 
small-$S^1$ and large-$S^1$ regimes. 
We also find concrete microscopic connections between  ${\cal N}=1$ and  ${\cal N}=2$   supersymmetric gauge theory dynamics and non-supersymmetric QCD dynamics. 
\end{abstract}


\maketitle

{\bf Introduction. }
The importance of spontaneous chiral symmetry breaking  ($\chi$SB)  in QCD 
for our understanding of nature is hard to overstate. For example, $\chi$SB 
gives the main contribution to the rest masses of the light mesons and 
baryons, and strongly constrains their interactions. However, the microscopic 
mechanism of $\chi$SB is still mysterious. The basic problem is that 
in the situations known to date in QCD, continuous $\chi$SB at zero matter density only happens at strong coupling  \footnote{At  high quark density in QCD,  a weak coupling version of
$\chi$SB  occurs via  color-flavor locking \cite{Alford:2007xm}.  }.
 So, while there are many phenomenological models of $\chi$SB, such as Nambu-Jona-Lasinio 
models \cite{Nambu:1961tp}, truncated Schwinger-Dyson models 
\cite{Roberts:1994dr}, and instanton liquid models \cite{Schafer:1996wv,Shuryak:2012aa,*Liu:2015jsa}, 
$\chi$SB happens outside of the regime where they are under systematic 
theoretical control.

Here we present a new mechanism of $\chi$SB, in a class of 4D non-Abelian 
gauge theories which is smoothly connected to standard 4D QCD with $N_c = 3$ 
and $N_f=3$. Our mechanism operates in a weakly-coupled regime which is under 
systematic theoretical control thanks to the technique of adiabatic 
continuity \cite{Unsal:2007vu,Unsal:2007jx,Unsal:2008ch,Shifman:2009tp,Shifman:2008ja,Argyres:2012ka,Poppitz:2012sw,Anber:2013doa,Dunne:2016nmc}. 
We are thus able 
to give a controlled microscopic derivation of the chiral Lagrangian of the 
Nambu-Goldstone (NG) bosons. Another historical mystery is the reason for the 
phenomenological successes of the naive quark model, built from ``constituent 
quarks" with masses of the same scale as the gauge-sector mass gap. We
show that constituent quark masses arise naturally from monopole-instantons.

{\bf The setting.} 
We consider $SU(N_c>2)$ gauge theory with a strong scale $\Lambda$ coupled 
to $N_f \le N_c$ flavors of fundamental Dirac fermions $\psi_{i}, i = 1, 
\cdots, N_f$ with a common mass $m_q \ll \Lambda$, as well as one  heavy  adjoint Dirac 
fermion $\lambda$ with mass $m_{\lambda} \gg 
\Lambda$.   On $\mathbb{R}^{1,3}$ the heavy adjoint fermion is a spectator field.  On $\mathbb{R}^{1,2} \times S^1$, it  plays a role in center stabilization, but otherwise it is still decoupled 
  from the dynamics of light states.  So for light states we deal with standard 
QCD if we set $N_c = 3$ and $N_f =2, 3$.   
The 
idea of adiabatic continuity is to find a way to put an asymptotically-free 
gauge theory on $\mathbb{R}^{1,2} \times S^1$ in such a way that its dependence 
on the spatial circle size $L$ is smooth.  If this condition is met, then one 
can get insight about the behavior of the theory for large 
$L$, where it is strongly coupled, by studying it for small $L$, where it is 
weakly coupled.  Satisfying the adiabaticity condition has been especially 
challenging\cite{Shifman:2009tp,Unsal:2008eg} in theories with continuous chiral symmetries, but in this paper 
we present a method to ensure it, allowing us to address chiral symmetry 
breaking.

{\bf Large L expectations.}  
If $m_q = 0$, the quantum theory has global 
symmetry  
\begin{align}
G= SU(N_f)_L \times SU(N_f)_R \times U(1)_Q \, . 
  \label{eq:non_ab}
\end{align}
Here, we already factored out  anomalous $U(1)_A$ which reduces to $ \mathbb{Z}_{2N_f}$ due to instanton effects. 
 It is believed 
that $G$ breaks spontaneously  to $SU(N_f)_V \times U(1)_Q$, and the 
low-energy dynamics of the resulting NG bosons are described 
by chiral perturbation theory
\begin{align}
\mathcal{L}_{\chi PT} = \frac{f_{\pi}^2}{4} \tr |\partial_{\mu} \Sigma|^2 
   - c \,\tr(M_q^{\dag} \Sigma +M_q \Sigma^{\dag}) + \cdots \, , 
\label{eq:chi_PT}
\end{align}
where $\Sigma = e^{i \Pi/f_{\pi}}$ and $M_q = m_q 1_{N_f}$ is the quark mass 
spurion field, transforming as $\Sigma \to L\, \Sigma R^{\dag}, M_q \to L\, 
M_q R^{\dag}$ under chiral rotations.  Equation~\eqref{eq:chi_PT} implies that the NG boson masses obey the famous Gell-Mann-Oakes-Renner \cite{GellMann:1968rz}  (GMOR) relation 
$f_{\pi}^2 m_{\pi}^2 = m_q \langle \bar{\psi} \psi\rangle$, where 
$\langle\bar{\psi} \psi\rangle =-2 c$. 

If generic imaginary 
chemical potentials $\mu$ are turned on for the Cartan subgroup $U(1)^{N_f-1}_L 
\times U(1)^{N_f-1}_R$ of   $G$
then $N_f-1$ NG 
bosons remain gapless, while the the rest pick up mass gaps $\sim |\mu|$ in 
the chiral limit $m_q = 0$.  So, for instance, for $N_f=2$, turning on an 
isospin chemical potential $\mu_I$ corresponds to shifting $\partial_{\mu} \Sigma 
\to \partial_{\mu} \Sigma +i [\mu_{I} \tau_3/2,\Sigma]$ \cite{Son:2000xc}.   With an imaginary $\mu_I$ the $\pi_{\pm}$ 
fields get positive mass gaps $\sim |\mu_I|$, while the $\pi_0$ remains gapless. 

{\bf Vacuum structure at small $L$.}  
To compactify the theory on $S^1$ we must choose boundary conditions for 
the fields.  As usual, we take the gluons $A_{M}, M=1,\ldots, 4$ to be 
periodic. 
For the fermions, there are many more choices: one can demand periodicity up 
to global symmetry transformations, see e.g.~\cite{Bedaque:2004kc,Sachrajda:2004mi,Kouno:2012zz,Kouno:2013mma,Kouno:2013zr,Poppitz:2013zqa,Misumi:2014raa,Iritani:2015ara}. This is equivalent to turning on expectation values for holonomies 
$\Omega_{\mathcal{F}} = {\rm P} e^{ i\int_{S^1} \mathcal{A}_4}$ of background flavor 
gauge fields $\mathcal{A}_{M}$ valued in the Lie algebra of 
Eq.~\eqref{eq:non_ab}, and can also be interpreted as turning on imaginary 
flavor chemical potentials $\mu \sim 1/L$.  Such twists clearly become 
irrelevant when $L \gg \Lambda$, but can become important when $L \ll 
\Lambda$.  

 Once $L\Lambda \ll 1$ the 4D gauge coupling at the scale $1/L$ becomes 
small, and the theory comes under semiclassical control, provided  that  the 
long-distance dynamics becomes Abelian.  Whether this happens depends on 
the expectation value for the gauge holonomy in the compact direction 
$\Omega = {\rm P} e^{i\int_{S^1} A_4} $.  At large $L$, we expect the 
``center-symmetric" result $\langle \tr \Omega^n \rangle \simeq 0, 
n = 1,\cdots, N_c-1$ due to strong eigenvalue fluctuations (despite 
the lack of an exact center symmetry unless $N_f = 0$ or $N_f = N_c$, see \cite{Kouno:2012zz}).  On the other hand, at small $L$ the eigenvalue fluctuations become small, and   $\langle 
\tr \Omega^n\rangle$ is determined by the minimum of the Wilsonian effective 
potential. 
The role of the massive $\lambda$ field is 
 to produce a $\mathbb{Z}_{N_c}$ center-symmetry-stabilizing 
contribution to the potential\cite{Unsal:2008ch,Myers:2007vc,Myers:2009df}, see also \cite{Hosotani:1983xw}, which can also be achieved by double-trace deformations \cite{Unsal:2008ch}.
The minimum of the potential is then attained with a $\Z_{N_c}$-center-symmetric configuration $\Omega  \sim  \textrm{diag}(1,\omega,\cdots, \omega^{N_c-1})$, 
where  $\omega = e^{2\pi i /N_c}$.

\begin{figure}[t]
\centering
\includegraphics[width=0.5\textwidth]{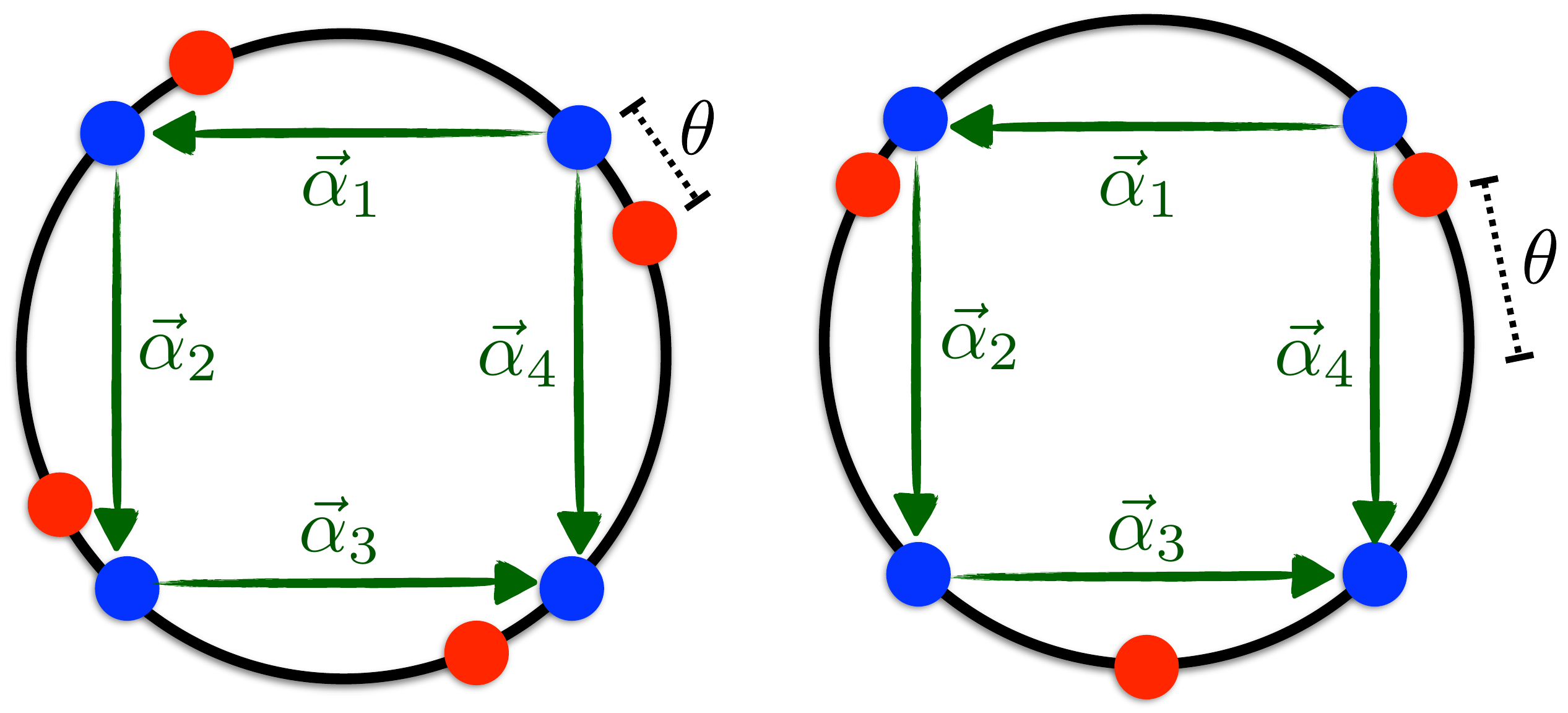}
\caption{Eigenvalue distributions for the gauge  holonomy $\Omega$ (blue dots) $N_c = 4$ and flavor holonomy $\Omega_F$ (red dots) with $N_f =4$ (on the left) and $N_f=3$ (on the right).  The $U(1)_Q$ holonomy $\Omega_Q = e^{i\theta}$ rotates $\Omega_F$ and $\Omega$ eigenvalues relative to each other.  The green arrows represent the four monopole-instanton events, which are labeled by the affine roots $\vec{\alpha}_{i}$.  Each time an arrow ``passes" a flavor holonomy eigenvalue, the associated monopole-instanton operator picks up two fermion zero modes.  So in the left $N_f=4$ figure, all monopole instantons have two zero modes, while in the right $N_f=3$ figure $\vec{\alpha}_{1}$ has no zero modes while $\vec{\alpha}_{2,3,4}$ each have two zero modes. }
\label{fig:holonomy_circle}
\end{figure}

Unlike the VEV of $\Omega$, which is determined dynamically, the flavor holonomy is an external parameter and can be chosen at will.    As emphasized in 
\cite{Dunne:2012ae,Dunne:2012zk,Cherman:2013yfa,Cherman:2014ofa,Dunne:2015ywa,Misumi:2014bsa,*Misumi:2014jua}, 
  some choices are better than others in the context of adiabatic continuity, and can lead to a smoother passage from large $L$ to small $L$.   
At the two extremes, we can choose a trivial flavor-holonomy $\Omega_F={\bf 1}_{N_f}$, or 
a non-trivial  flavor-holonomy $\Omega_F 
\sim \textrm{diag}
(1,\omega_F,\cdots, \omega_F^{N_f-1})$, where 
$\omega_F = e^{2\pi i /N_f}$, which preserves a $\mathbb{Z}_{N_f}$ flavor-center symmetry.  
Center-symmetric  gauge and flavor holonomy configurations are  
illustrated in Fig.~\ref{fig:holonomy_circle}.  Choosing $\Omega_F$ to be  $\mathbb{Z}_{N_f}$ symmetric has rather dramatic implications, as discussed below.  

At small $L$, $A_4$ acts as an compact adjoint Higgs field. At long distances 
the center-symmetric VEV for $\Omega$ leads to an Abelianization of the 
dynamics:
\begin{align}
    SU(N_c) \rightarrow  U(1)^{(N_c-1)} \,.
    \label{eq:higgsing}
\end{align}
This is similar to what happens in the 3D Polyakov model \cite{Polyakov:1976fu} 
and Seiberg-Witten analysis of $\mathcal{N}=1$ and $\mathcal{N}=2$ super-Yang-Mills (SYM) theories \cite{Seiberg:1994rs,*Seiberg:1994aj}, but the fact 
that $A_4$ is a compact variable leads to important differences in the physics.
The Abelianization happens at the scale of the lightest $W$-boson mass $m_W= \frac{2\pi}{L N_c}$.

{\bf Perturbative small L physics.}  
In perturbation theory the Abelianized small $L$ regime can be described by 
a 3D effective field theory.  The lightest fields are the Cartan gluons 
$A^{i}_{\mu}$, where $i = 1, \ldots, N_c-1$ and from now on $\mu = 1,2,3$.  In 
fact  $A^{i}_{\mu}$ are massless to all orders in perturbation theory. This 
is easiest to see by using an Abelian duality transformation $F^{i}_{\mu \nu} 
=  g^2/(2\pi L) \epsilon_{\mu\nu\alpha} \partial^{\alpha} \sigma^{i}$ relating 
the Cartan gluons to 3D scalars $\sigma^{i}$, which we call ``dual photons".  
If we write $\sigma_i = \vec{\alpha}_i \cdot \vec{\sigma}$, where 
$\vec{\alpha}_i$ are the simple roots of the algebra $\mathfrak{su}(N_c)$ 
and $\vec{\cdot}$ denotes an $N_c$\,-vector, then the dual photons enjoy 
an emergent topological shift symmetry
\begin{align}
 [U(1)_{ J}]^{N_c -1}: 
  \vec \sigma  \rightarrow \vec \sigma   + \vec \epsilon,  \qquad 
  \vec {\cal J}_{\mu} = \partial_{\mu} \vec \sigma  
\label{shift}
\end{align}
in perturbation theory.  The Noether currents $\vec {\mathcal{J}}_{\mu}$ can be 
identified with  the (Euclidean) magnetic field $\vec{B}_{\mu} = 
\epsilon_{\mu \nu \alpha}\vec{F}^{\nu \alpha}$ using  the Abelian duality relation, 
and current conservation is just the statement of the absence of magnetic 
monopoles $\partial_{\mu} \vec {\cal J}_{\mu} =\partial_{\mu} \vec{B}_{\mu}   = 0$. 
Therefore, to all orders in perturbation theory, the
dual photons must remain gapless.  Their action looks like
\begin{align}
S_{\sigma} = \int d^3{x}\, \frac{g^2}{8\pi^2 L}(\partial_{\mu} \vec{\sigma})^2.
\label{eq:dual_photon_action}
\end{align}
Quarks are charged under the Cartan subgroup of $SU(N_c)$, so one might 
be concerned that the resulting 3D QED theory, with gauge coupling $g_3^2 = 
g_{\rm YM}^2(m_W)/L$, might flow to strong coupling in the far infrared if 
$m_q = 0$.  However, this does not happen for generic values of the $U(1)_Q$ 
holonomy $\Omega_Q = e^{i \theta}$.  The long-distance perturbative action for 
the quarks is
\begin{align}
\!\!\!S_{\psi} = L\int d^{3}x \, \bigg\{ 
 & \bar{\psi}_L \left[ \gamma^{\mu}D_{\mu}  
    + \gamma^4(A_{4} +\mathcal{A}_{4} + \theta/L)\right]\psi_L  \\
 & + m_q \bar{\psi}_L \psi_R + L \leftrightarrow R \nonumber \}\, . 
\end{align}
So even if $m_q=0$, the center-symmetric color and flavor holonomies give 
$\psi_{n,a}$ a chirally-invariant gap $\frac{2\pi}{L}(\frac{n}{N_c} 
+ \frac{a}{N_f} +\theta )$. So long as $\theta  \gtrsim g_3^2 L = g_{\rm YM}^2$, within perturbation theory all of the quarks decouple 
from the Cartan gluons at long distances, and there is no flow to strong coupling.

{\bf Monopole-instantons.}    
Due to Eq.~\eqref{eq:higgsing} our theory has $N_c$ types of monopole-instantons
\cite{Lee:1997vp,Kraan:1998sn,Davies:1999uw,Unsal:2008ch}. In the 
center-symmetric background, they all have identical Euclidean actions 
$S_0 = \frac{S_{\cal I}}{N_c}= \frac{8\pi^2}{g^2N_c}$, where $S_{\cal I} = 
\frac{8 \pi^2}{g^2}$ is the 4d  instanton action. In this setting the 
4d  instanton is  a composite configuration built from the $N_c$ 
monopole-instantons. These solutions are associated with the affine root 
system of $\frak{su}(N_c)$ Lie algebra: $N_c-1$ correspond to the simple 
roots, while the remaining one is associated with the affine root, and is 
due to the compact nature of the adjoint Higgs field 
\cite{Lee:1997vp,Kraan:1998sn}.  
 
 To understand the contributions of these finite-action field configurations, 
we note that monopole-instantons carry two types of topological quantum 
numbers, magnetic and topological charge $(Q_m, Q_T) = \left( \frac{g}{4\pi} 
\int_{S_2} {F \cdot dS}, \frac{1}{16 \pi^2} \int {\tr F_{\mu \nu} \widetilde 
F^{\mu \nu}} \right) 
$, given by $(Q_m, Q_T) = 
\pm \left( \alpha_i, 
1/N_c  \right)$ for monopoles and anti-monopoles.
 
These field 
configurations couple to the dual photons, and, if $N_f = 0$ --- that is, 
in pure Yang-Mills theory deformed by a heavy adjoint fermion --- the 
associated amplitudes take the form
\begin{align}
\mathcal{M}_{i} = e^{-S_0} 
        e^{   i  \vec\alpha_i  \cdot \vec\sigma  }, \; \textrm{no fermions}\, .
\label{eq:monopoles} 
\end{align}
The usual 4d  instanton amplitude is 
\begin{align}
\label{eq:instanton}
{\cal I}_{4d} \sim \prod_{i=1}^{N_c}{\cal M}_{i}  \sim  e^{-\frac{8 \pi^2}{g^2}}, 
 \; \textrm{no fermions}\, . 
\end{align}
The 4d instanton has vanishing magnetic charge, and does not 
couple to dual photons. Consequently, it does not play any important role 
in the long-distance dynamics, which is determined by the monopole-instantons. 
The fact that the monopole-instantons carry magnetic charge means that, for 
$N_f=0$, their proliferation in the vacuum leads to an explicit non-perturbative 
breaking of the topological $[U(1)_{ J}]^{N_c -1}$ shift symmetry of the $\sigma_{i}$ fields, 
$\partial_{\mu} \vec {\cal J}_{\mu} =\partial_{\mu} \vec{B}_{\mu}   = \vec \rho_m$, where 
$ \vec \rho_m$ is the (non-perturbative)  magnetic charge density for the monopole-instantons. 
The monopoles and anti-monopoles generate a potential $V(\vec{\sigma}) \sim -
m_W^3 e^{-S_0} \sum_{i} \cos(\vec\alpha_i  \cdot \vec\sigma)$, and fluctuations
around its minimum $\vec{\sigma} = 0$ have mass $m_{\sigma}^2 \sim m_W^2 
e^{-S_0}$, as discussed in Ref.~\cite{Unsal:2008ch}, so that the deformed YM 
theory develops a non-perturbative mass gap.   The long-distance $N_f=0$ 
theory exhibits both confinement of electric charge with finite string 
tension and a non-perturbative mass gap for all gauge fluctuations. 

{\bf Fermion zero modes.}   
When $m_q = 0$,  
the chiral anomaly implies the presence 
of $2N_f$ fermion zero modes in the  instanton background.  This modifies 
the  instanton amplitude from Eq.~\eqref{eq:instanton} to 
\cite{tHooft:1976fv}
\begin{align}
\label{eq:instanton_fermions}
\mathcal{I}_{4d}  &\sim   e^{- \frac{8 \pi^2}{g^2}} 
  \,\, {\rm det}_{a,b} \left[ \bar{\psi}_{L,a} \psi_{R,b}  \right]   
\end{align}  
which is invariant under $\Z_{2N_f}$ but not $U(1)_A$.  However, the 't Hooft 
vertex in Eq.~\eqref{eq:instanton_fermions} is invariant under $G$ from Eq.~\eqref{eq:non_ab}
because these symmetries are not anomalous, 
 and must  be preserved in all effective interaction vertices. 

At small $S^1$, the 4d instanton splits into monopole-instantons, and if 
$\Omega_F  \sim  1_{N_f} $ and $\psi_j(x_4+L)= e^{i\theta}\psi_j(x_4)$,   
 the Nye-Singer index theorem implies that  
all the fermion zero modes localize on a single monopole \cite{Nye:2000eg,Poppitz:2008hr}, say, 
${\cal M}_{1}$. Then the monopole amplitudes are given by \begin{align}
\label{eq:monopoles_fermions}
{\cal M}_{1} &= e^{-S_0}    e^{ i  \vec\alpha_1  \cdot \vec\sigma }  
 \,\, {\rm det}_{a,b} \left[ \bar{\psi}_{L,a} \psi_{R,b}  \right] \cr 
{\cal M}_{i} &=    e^{-S_0} e^{ i  \vec\alpha_i  \cdot \vec\sigma  }, \qquad i=2, \ldots N_c
\end{align}
All $\mathcal{M}_i$ are invariant under $G$ from Eq.~\eqref{eq:non_ab}, 
and $\prod_i \mathcal{M}_i$ is equivalent to Eq.~\eqref{eq:instanton_fermions}.
With the choice $\Omega_F \sim 1_{N_f}$ there is no $\chi$SB at weak coupling.  
The reason  is  that all $2N_f$ fermi zero modes are  localized at one monopole-instanton, 
and the semi-classical weight $e^{-S_0}$ is not large enough to break chiral symmetry.  
In this case,  there must be a chiral phase transition between small $L$ and large $L$ at 
the scale $L = \Lambda^{-1}/N_c$, as explained in \cite{Unsal:2008eg}.

\begin{figure}[t]
\centering
\includegraphics[width=0.40\textwidth]{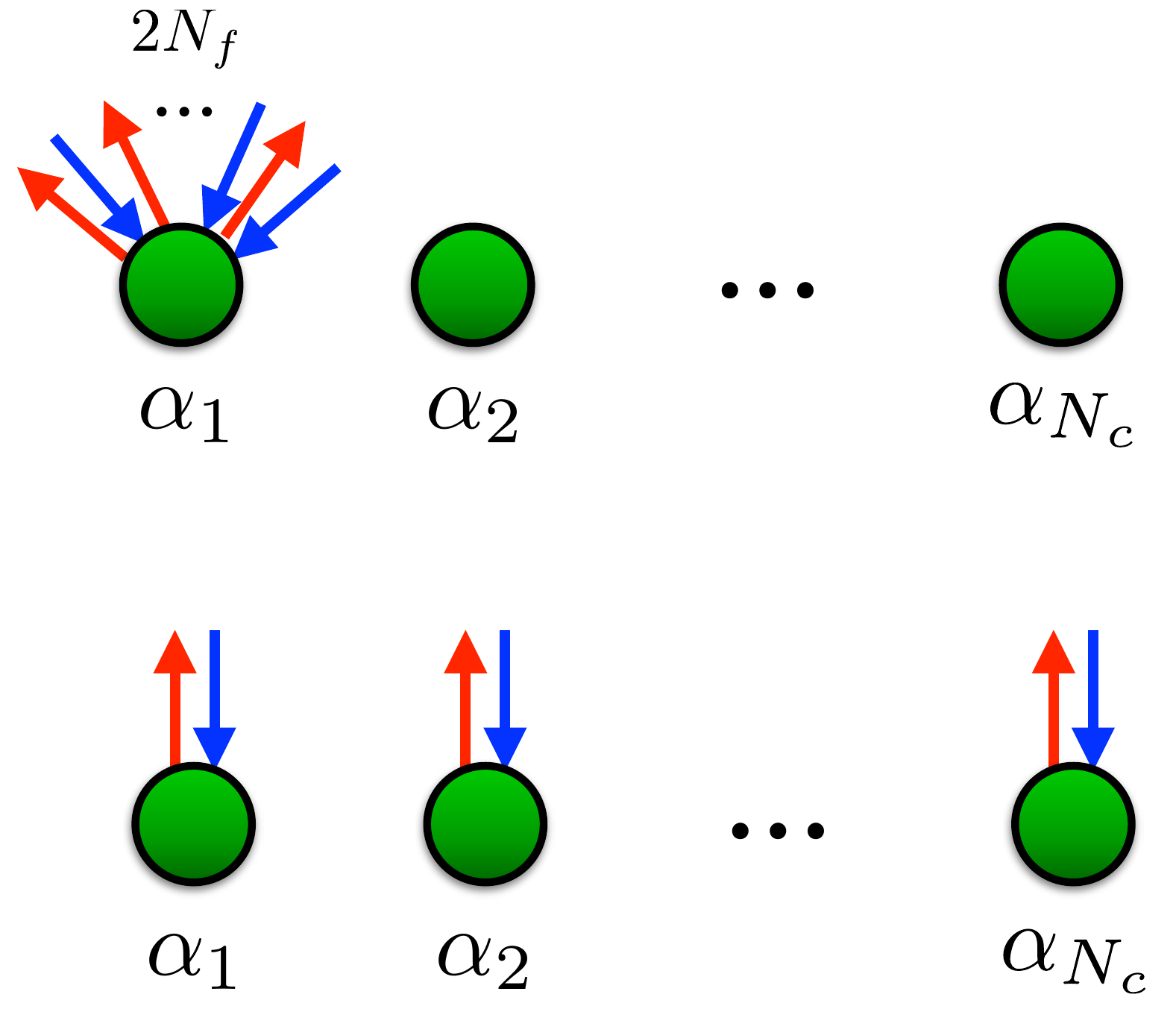}
\caption{The distribution of fundamental fermion zero modes for the 
monopole-instantons with a trivial flavor holonomy (top row) and a 
$\mathbb{Z}_{N_f}$-symmetric flavor holonomy (bottom row) for $N_f = N_c$. 
The latter distribution is identical  to  ${\cal N}=1 $ SYM on  $\R^3 \times S^1$ in which monopole-instantons saturate the chiral condensate. 
}
\label{fig:zero-modes}
\end{figure}

As the $U(1)_Q$ holonomy is dialed, the
$2N_f$ zero modes {\it jump collectively}  from one monopole instanton 
to the next  at  discrete values of  $\theta$, e.g, at  
 $\theta = \frac{2\pi k}{N_c}, k \in \mathbb{Z}$ for $N_f=N_c$ 
\cite{GarciaPerez:1999ux, Bruckmann:2003ag}.  In other words, the index is only a piece-wise constant  function, and exhibits jumps  with respect to $\theta$. 
This is interpreted as wall-crossing in \cite{Moore:2014gua}.
 
{\bf Fractional jumping.} The main new idea of this work is a refinement of collective jumping.  
If we dial the $\Omega_F$ holonomy from the trivial configuration $\Omega_F = 1_{N_f}$ toward a configuration that  respects  the flavor center symmetry $\mathbb{Z}_{N_f}$, 
this shifts the frequency quantization of different flavor  components, and leads to  {\it fractional  jumping} (in units of two) of the $2N_f$ fermionic zero modes onto the  
$N_c$  monopole-instantons as evenly as possible.  This fractional jumping phenomenon, as well as the collective jumping, is illustrated in Fig.~\ref{fig:holonomy_circle}. 
At long distances the twisted  boundary conditions result in an explicit breaking of the global symmetry $G$ down to its maximal Abelian subgroup
\begin{align}
 G_{\rm max-ab} = [U(1)_V \times U(1)_A]^{N_f-1} \times U(1)_Q \,.
\label{max-ab}
\end{align} 
 With the twist, $N_f$ of the monopole operators carry two fermi zero modes each, while the other 
$N_c -N_f$ monopole-instantons have no fermion zero modes.  The monopole amplitudes are, schematically, 
\begin{align}
\label{monopoles-2}
{\cal M}_{i} &=    e^{-S_0}  e^{i \vec\alpha_i \cdot \vec\sigma } 
      (\bar{\psi}_{L,i} \psi_{R,i}), \qquad i=1, \ldots N_f\, ,  \\
{\cal M}_{k} &=  e^{-S_0}   e^{i \vec\alpha_k \cdot \vec\sigma }, 
                                   \qquad k=N_f+1, \ldots N_c \, . 
\end{align}
and are illustrated in Fig.~\ref{fig:zero-modes}. The  individual zero mode vertices $(\bar{\psi}_{L,i} \psi_{R,i})$ 
transform non-trivially under  $[U(1)_A]^{N_f-1}$,   and naively this could lead one to the puzzling conclusion that there is an anomaly for the corresponding 
$U(1)_A$ factors. But  this is impossible, because  $G_{\rm max-ab}$ is a subgroup of a non-abelian chiral symmetry, which is non-anomalous. 

The resolution of this puzzle comes from an axial-topological symmetry intertwining effect, also seen in Ref.~\cite{Poppitz:2009tw}.  The
$[U(1)_A]^{N_f-1}$ symmetry intertwines with $[U(1)_{ J}]^{N_f -1}$ subgroup of  Eq.~\eqref{shift}, such that the symmetry phase of the fermion bilinear is cancelled exactly 
by a shift of the  dual photon field, $\vec{\sigma}$: 
 \begin{align}
\begin{array}{ccc}
(\bar{\psi}_{L,k} \psi_{R,k})    & \rightarrow &
   e^{i \epsilon_k } (\bar{\psi}_{L,k} \psi_{R,k}),  \\[0.2cm]
e^{i \vec\alpha_k  \cdot \vec\sigma  } &\rightarrow & 
   e^{-i \epsilon_k }    e^{i\vec\alpha_k  \cdot \vec\sigma} .
\end{array} 
\end{align} 
 In this way, the monopole-instanton amplitude is invariant under the expected symmetries.

Since the $N_f$ parameters $\epsilon_k$ must satisfy $\sum_{k=1}^{N_f} \epsilon_k
= 0$ they can be written as $\epsilon_k =  \bm{\alpha}_k \cdot \bm{\epsilon}$ 
where $\bm{\alpha}_k$ are the simple roots of $\mathfrak{su}(N_f)$ and bold 
symbols denote $N_f$-vectors.  Writing $\vec{\sigma} = (\bm{\sigma}, 
\tilde{\sigma})$, where $\tilde{\sigma}$ is an $N_c-N_f$ vector, we find 
that $\bm{\sigma}$ enjoys an \emph{unbroken} shift symmetry,
\begin{align}
 \bm{\sigma}   \rightarrow  \bm{\sigma}   +  \bm{\epsilon}  \, ,
\label{eq:trans}
\end{align}
while the $[U(1)_J]^{N_c-N_f}$ shift symmetry of $\tilde{\sigma}$ is  {\it explicitly} broken.

{\bf Chiral symmetry breaking.}  
Now consider the fluctuations of $\bm{\sigma}$ around any point on its vacuum 
manifold $U(1)^{N_f-1}$,  along with the fluctuations of $\tilde{\sigma}$ around  the bottom of its potential at $\tilde{\sigma}=0$. 

The choice of a point  on the vacuum manifold {\it spontaneously} breaks the $[U(1)_{A}]^{N_f-1}$. Then Eq.~\eqref{eq:trans} 
implies that  $N_f-1$  dual photons enjoy a shift symmetry and remain \emph{gapless} non-perturbatively. 
Another, more microscopic, way to see this is that there is no analog of the magnetic bion mechanism of mass generation from adjoint QCD in the present context, because all magnetically-charged topological molecules which couple to $\bm{\sigma}$ have uncompensated fermion zero modes. The remaining 
$N_c-N_f$ dual photons $\tilde{\sigma}$ pick up masses $m_{\sigma} \sim m_W e^{-S_0/2}$, as in pure YM theory.  

It is also worth noting that in ${\cal N}=2$ SYM theory compactified on $\R^3 \times S^1$ \cite{Seiberg:1996nz}, the gaplessness of the dual photon is due to an
identical mechanism of axial-topological  symmetry intertwining.  In Ref.~\cite{Seiberg:1996nz}, however, supersymmetry implies that an entire ${\cal N}=2$ multiplet remains gapless, and there is a quantum moduli space associated to the elementary scalar fields as well.   
In our context, only the dual photons are protected, and there are no elementary scalars. Nevertheless, it is intriguing to see such close parallels between a theory with extended supersymmetry and QCD.

 The basic physical phenomenon is that gapless dual photons are transmuted 
into the NG bosons of the spontaneously broken Abelian chiral 
symmetry. This is the chiral symmetry group that remains exact with our choice 
of twisted boundary conditions. Indeed, if we define $\Pi^{\prime} =  \sum_{a=1}^{N_f} \pi_a T_a$, where $T_a$ are the Cartan generators of $\mathfrak{su}(N_f)$, $\pi_a = g/(2\pi L) \sigma_a$, and $\Sigma^{\prime} = e^{i \Pi^{\prime}/f_{\pi}}$, 
then the action for $\bm{\sigma}$ can 
be written as 
\begin{align}
\!\!\!\!\!\!\!S_{\bm{\sigma}} =  L\!\! \int d^{3}x \left[ \frac{f_{\pi}^2}{4} 
  \tr \partial_{\mu} \Sigma^{\prime} \partial^{\mu} \Sigma^{\prime \dag}   - c \,\tr(M_q^{\dag} \Sigma' +\textrm{h.c.})\right]\!,
\label{eq:small_L_chiPT}
\end{align}
with the derivation of the $c$ term given below.  This precisely matches the dimensional reduction of  Eq.~\eqref{eq:chi_PT} 
to 3D, with $\Sigma'$ interpreted as the restriction of the  
$SU(N_f)$-valued chiral field $\Sigma$ to its maximal torus, which parametrizes
the exactly massless NG bosons. In the regime of small $L$ 
the value of $f_{\pi}$ can be calculated by gauging $G_{\rm max-ab}$,
\begin{align}
f^2_{\pi} = \left(\frac{g}{\pi L \sqrt{6}}\right)^2 = \frac{N_c \, \lambda\, m_W^2 }{24\pi^4} \, ,
\end{align}
where $\lambda = g^2 N_c$.  This implies that starting from the weak-coupling, small-$L$, side of 
compactified QCD we have been able to derive the chiral Lagrangian expected 
from the strong coupling, large-$L$ side.  This supports the expectation 
that the compactification is indeed adiabatic, with no phase 
transitions as a function of $L$. 

{\bf Constituent quark masses and the chiral condensate.}  
We now work out the effect of turning on light quark masses, as well as 
explain the origin of the ``constituent quark" masses. First, consider 
turning on a small ``current" quark mass $m_q$. This corresponds to a term 
in the Lagrangian of the form ${\cal L}_m=m_q(\bar\psi_L\psi_R+\mathbf{h.c.})$. In order
to obtain the contribution to the effective action we saturate the integral
over the fermion zero mode with a single mass insertion. We find
\begin{align}
{\cal L}_m = m_W^2 m_q e^{-S_0} e^{i \bm{\sigma}_k \cdot \bm{\sigma}} + \mathrm{h.c.}\, ,
\end{align}
which leads to the $c$ term in Eq.~\eqref{eq:small_L_chiPT}.  The VEV of the monopole operator determines the vacuum 
energy density ${\cal E}=-c\,\tr [M_{q}+h.c.]$ and the chiral condensate. 
We get $\langle\bar\psi\psi\rangle = -2 m_W^{-3}e^{-S_0}$.  The GMOR relation $f_{\pi}^2 m_{\pi}^2 = m_q \langle \bar{\psi}\psi \rangle$ is of course also reproduced.

We note that if $N_f=N_c$ there are some remarkable relations to ${\cal N}=1$ SYM.  In both theories, all $N_c$ monopole-instantons  acquire 2 fermion zero modes, the leading order beta 
function is $b=3N_c$ (on scales below $m_{\lambda}$), and the chiral condensate is saturated by the monopole-instantons.  These observations may be the microscopic 
explanation of   why the chiral condensate in  ${\cal N}=1$ SYM  is so close to the one in real QCD \cite{Armoni:2014ywa}.


Next, consider the notion of a constituent quark mass.  For vanishing 
$m_q$ the long-distance fermion effective Lagrangian includes the term
$\sum_{k=1}^{N_f}( m_W e^{-S_0} 
  e^{i \bm{\alpha}_k \cdot \bm{\sigma}} 
\bar{\psi}_{L,k} \psi_{R,k} + \mathrm{h.c.} )$.
So once the magnetic flux part $e^{i \bm{\alpha}_k \cdot \bm{\sigma}}$ of the 
monopole-instanton operator acquires a  VEV, the lightest quark modes 
pick up a non-perturbative chiral-symmetry breaking ``constituent quark" 
mass $m_{\rm constituent} \sim m_W e^{-S_0}$.  


{\bf Outlook. }  
We have described a new non-perturbative mechanism for $\chi$SB in 4D QCD 
which operates at weak coupling, and hence is under full theoretical control.  
The calculation is based on  adiabatic compactification, and a key 
ingredient is the use of a $\mathbb{Z}_{N_f}$-symmetric flavor holonomy.  
Spontaneous
chiral symmetry breaking arises from monopole-instanton vertices by 
color-flavor transmutation, and the NG pions originate from 
massless dual photons. This mechanism is somewhat reminiscent of chiral
symmetry breaking via color-flavor locking in high density QCD.  We also emphasize that our mechanism operates within the regime where monopole-instanton gas is dilute, which is a major difference from phenomenological instanton liquid models.
 We provided a microscopic derivation of the chiral Lagrangian. The
structure of the Lagrangian supports the idea that the small $S^1$ 
regime and the $\mathbb{R}^4$ limit are continuously connected.   

 Our results open a new playground for the exploration of 4D non-supersymmetric gauge theory dynamics 
 in a fully calculable setting with confinement and chiral symmetry breaking.   Some of the issues which are ripe for exploration include generalizations to $N_f > N_c$, to other gauge groups, $\chi$SB in chiral gauge theories, explorations of the  spectrum of non-Goldstone excitations, and many more.
 
{\bf Acknowledgments.}  We are grateful to P.~Draper, T.~Sulejmanpasic, and M.~Wagman for discussions.  This work 
was supported in part by the U. S. Department of Energy under the grants DE-FG02-00ER-41132 (A.~C.) and DE-FG02-03ER41260 (T.~S. and M.~U.).

\bibliographystyle{apsrev4-1}
\bibliography{small_circle} 
\end{document}